\documentclass[10pt,twocolumn]{article}

\usepackage{graphicx}
\usepackage[english]{babel}

\usepackage{amsmath}

\newlength{\plotwidth}
\allowdisplaybreaks[2]

\newcommand{\gev}{\operatorname{GeV}}

\newcommand{\lsim}{\smash{\raisebox{-4pt}{%
    $\,\stackrel{\textstyle <}{\sim}\,$}}}
\newcommand{\gsim}{\smash{\raisebox{-4pt}{%
    $\,\stackrel{\textstyle >}{\sim}\,$}}}

\begin{document}


\title{\raggedleft{\normalsize DESY 07-194} \\[0.5em]
  \centering\textbf{\LARGE Implications of HERA measurements for
    LHC}\,\footnote{%
    Talk presented at the XXIII International Symposium on Lepton and
    Photon Interactions at High Energy (LP07), Daegu, Korea, 13--18
    August 2007}}

\author{M. Diehl\\
\textit{Deutsches Elektronen-Synchroton DESY, 22603 Hamburg, Germany}}

\date{\parbox{0.9\textwidth}{\small%
  \textbf{Abstract:} 
  I discuss the theoretical understanding of key measurements at HERA
  and their relevance for physics at LHC, focusing on recent
  developments for structure functions and for diffraction.
}}

\maketitle


\section{Introduction}
\label{sec:intro}

In this talk I discuss key measurements at HERA and their impact on
physics at LHC, focusing on the topics of structure functions and of
diffraction.  Many important features of the final state such as jets,
multiple interactions, etc.\ will not be covered for lack of time.  A
wealth of up-to-date information can be found in the presentations of
the Workshop on HERA and the LHC \cite{hera-lhc}.


\section{Structure functions}
\label{sec:struct}

\begin{figure*}
\begin{center}
\includegraphics[width=0.73\textwidth,%
  bb=0 27 530 530,clip=true]{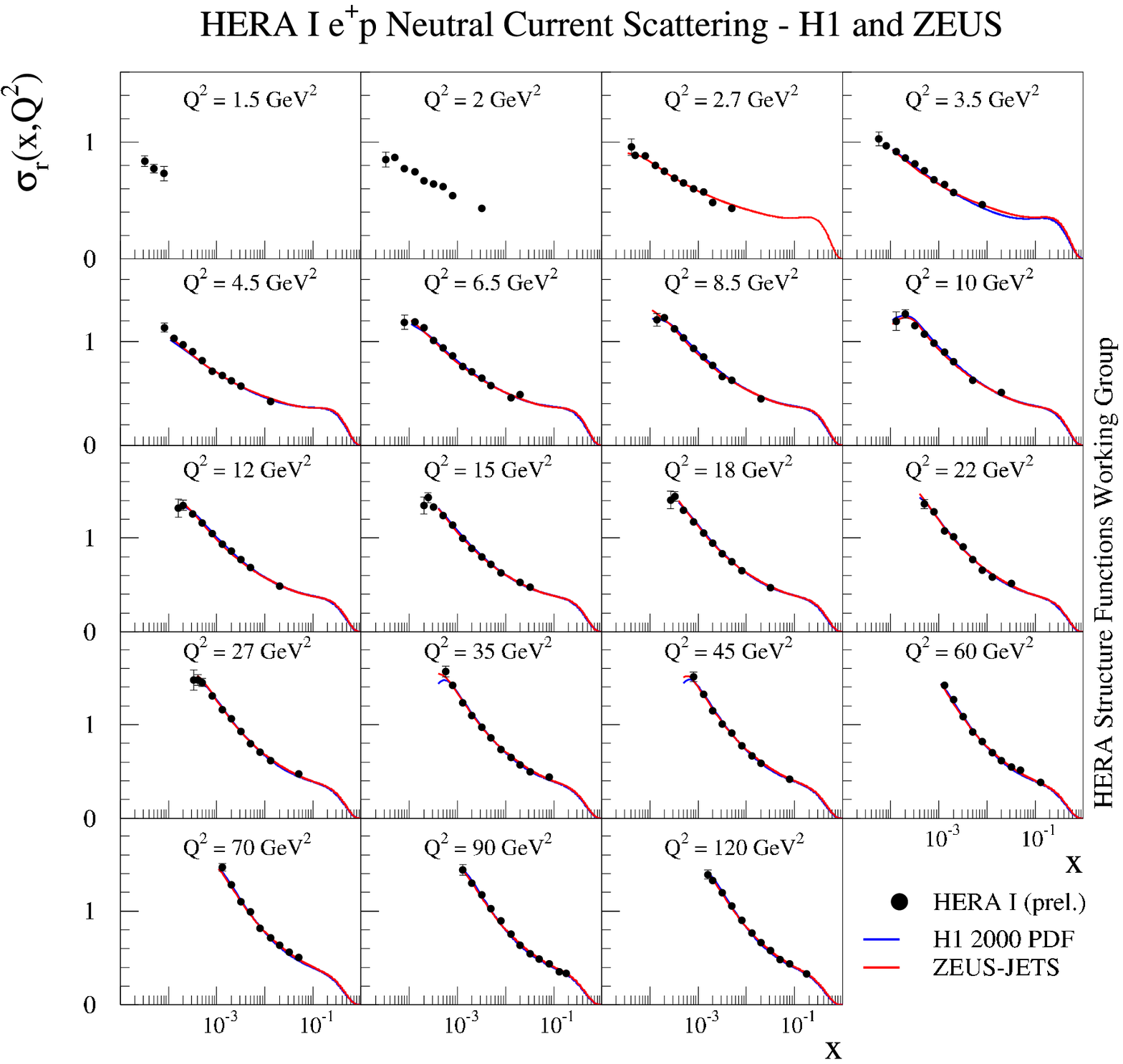}
\includegraphics[width=0.73\textwidth,clip=true,%
  bb=0 27 530 505,clip=true]{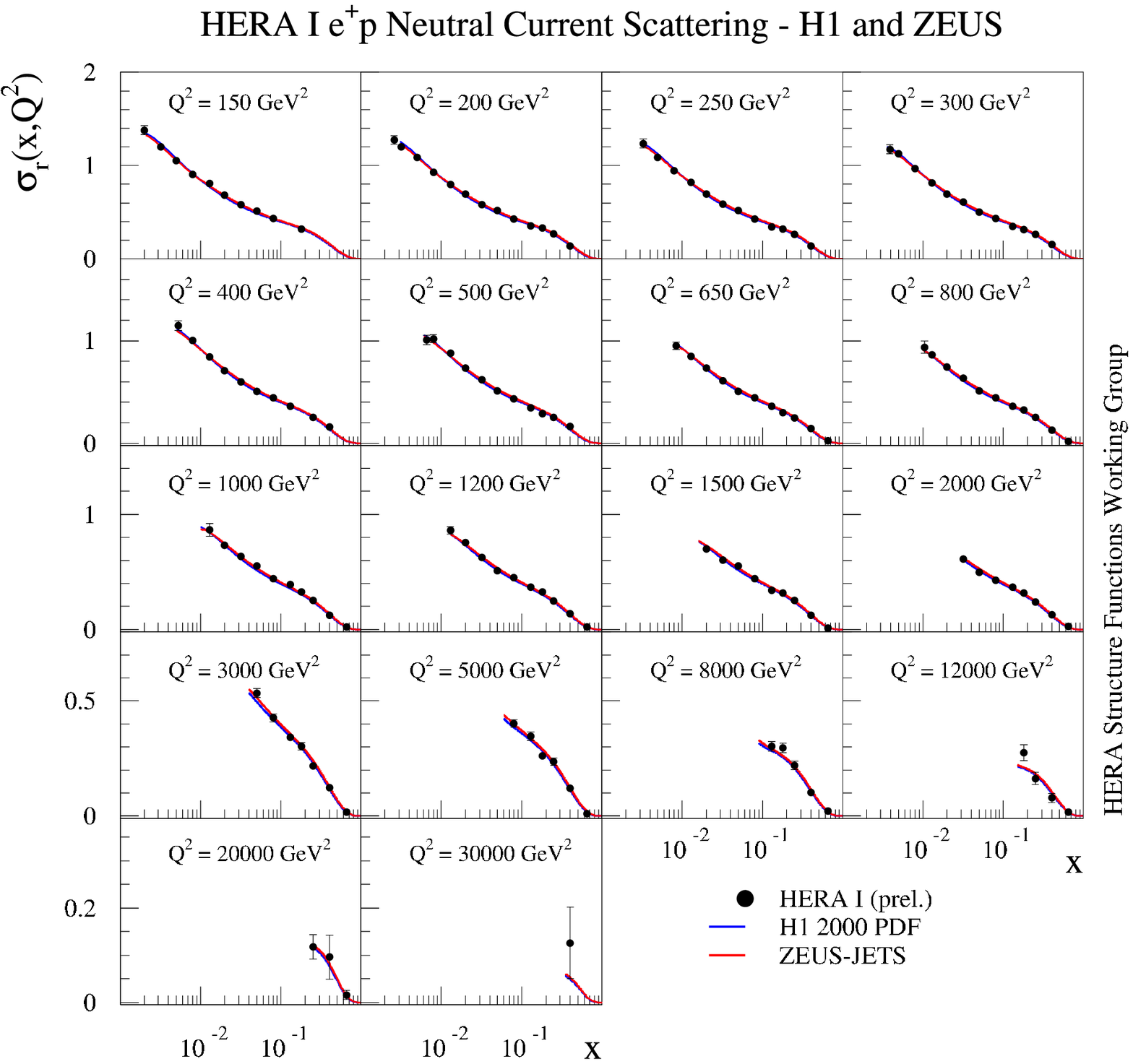}
\end{center}
\caption{\label{fig:hera-f2} HERA data for inclusive deep inelastic
  scattering \protect\cite{h1-zeus}.  When neglecting $Z$ exchange and
  the longitudinal structure function $F_L$, the reduced cross section
  is $\sigma_r \approx F_2(x,Q^2)$.}
\end{figure*}

A fundamental observable in deep inelastic scattering (DIS) is the
structure function $F_2$ of the proton.  A chief discovery of HERA is
its rapid rise with increasing energy or decreasing Bjorken-$x$, shown
in Fig.~\ref{fig:hera-f2}.  This has triggered many theory
developments, which continue to provide insight into the dynamics of
QCD at high energies.  In addition, the wide kinematic range and high
precision of the HERA structure function data makes them a key input
to the determination of parton densities (PDFs), which are in
particular needed to calculate the effective luminosities of colliding
partons at LHC.  The increasing precision of HERA data has been
matched by theoretical calculations up to next-to-next-to-leading
order (NNLO) in $\alpha_s$ \cite{Moch:2004pa}.

\begin{figure*}
\parbox{0.65\textwidth}{
\includegraphics[width=0.6\textwidth]{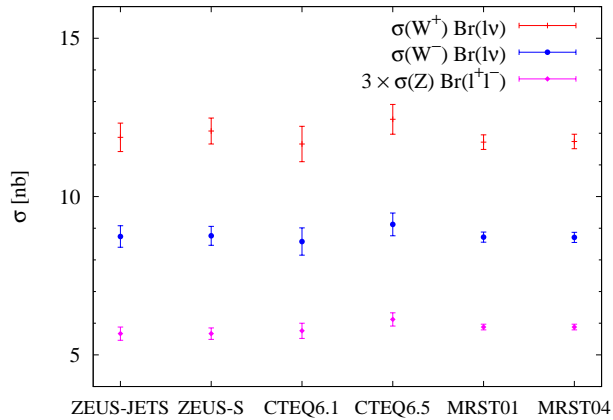}
}
\parbox{0.34\textwidth}{ 
\caption{\label{fig:xsec} Cross sections for $W$ and $Z$ production
  with leptonic decay at LHC, calculated with different sets of PDFs
  and their error estimates.  The numbers have been taken from Table~1
  in \protect\cite{CooperSarkar:2007pj} and correspond to the full
  rapidity range of the produced gauge boson.}
}
\end{figure*}

\begin{figure}
\begin{center}
\includegraphics[width=0.5\textwidth]{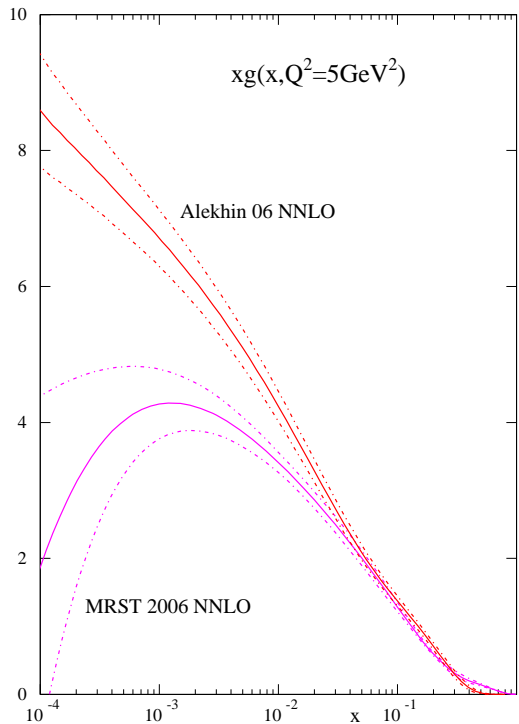}
\end{center}
\caption{\label{fig:glu} Gluon distributions including error bands
  from two recent analyses
  \protect\cite{Martin:2007bv,Alekhin:2006zm}.  The figure is taken
  from \protect\cite{Martin:2007bv}.}
\end{figure}

An example of a process where high precision is needed at LHC is the
production of a weak gauge boson, $W$ or $Z$, which has been proposed
as a possible luminosity monitor.  Fig.~\ref{fig:xsec} shows the
corresponding cross sections calculated with different recent sets of
PDFs.  The error bars on the cross sections reflect the errors
provided in the PDF parameterizations.  It is important to keep in
mind that they quantify the errors on the data to which the PDFs have
been fitted, but not uncertainties due to the choice of
parameterization, details of the theory description, or the data
selection.  As is clear from the figure, one can be misled when taking
the errors in a given PDF parameterization as a measure for the actual
uncertainty on the parton densities.  The same conclusion has been
obtained in an earlier study of Higgs production
\cite{Djouadi:2003jg}, which is one of the prime signal channels at
LHC.
An illustration of how much even the most recent PDF extractions can
differ is given in Fig.~\ref{fig:glu}, which shows the gluon
distribution obtained in the studies \cite{Martin:2007bv} and
\cite{Alekhin:2006zm}, both of which describe DIS at NNLO.  The shape
of the distribution is drastically different in the two analyses, and
below $x=10^{-2}$ their respective error bands do not overlap.

\subsection{Heavy flavor production}

The most important observable for determining the gluon density at
small $x$ is the scaling violation in the structure function
$F_2(x,Q^2)$.  Because of cross talk between gluons and sea quarks in
the $Q^2$ evolution, this leaves however room for ambiguities such as
the one in Fig.~\ref{fig:glu}.  An observable that may provide
independent constraints is the structure function $F_2^{c}$ for
inclusive charm production.  Several schemes can be used to describe
charm production in $ep$ or $pp$ collisions:
\begin{itemize}
\item The fixed flavor number scheme (FFNS) treats only the light
  quark flavors $u$, $d$, $s$ as partons and calculates the production
  of charm from the splitting $g\to c\bar{c}$ at fixed order in
  $\alpha_s$.  Most suitable when the hard scale $Q^2$ in the process
  is comparable to $m_c^2$, this scheme turns out to work rather well
  in most of the relevant kinematics for $F_2^c$ at HERA.  At very
  high scales, it is bound to fail because it misses large logarithms
  $\alpha_s^n \log^n (Q^2/m_c^2)$ from higher orders.
\item In the zero-mass variable flavor number scheme (ZM-VFNS), charm
  is treated as a massless parton, so that the logarithms just
  mentioned are resummed to all orders by the evolution of the parton
  densities.  Such a scheme is adequate for charm production at high
  transverse momenta at HERA, Tevatron, and especially at LHC, where
  important signal channels contain charm in the final state.
  Neglecting the charm quark mass, this scheme is however inadequate
  for describing data with $Q^2 \sim m_c^2$.
\item General-mass variable flavor number schemes (GM-VFNS) have been
  devised to interpolate smoothly between the two extremes just
  described.  Progress has recently been made on nontrivial issues
  that arise when consistently matching the descriptions with $n_f$
  and $n_f+1$ light quark flavors, see e.g.\
  \cite{Thorne:2006qt,Tung:2006tb}.
\end{itemize}
An analogous discussion holds for the production of bottom quarks.
Here again, it is only an interpolating scheme that can simultaneously
use input from the bottom structure function $F_2^b$ at HERA and
provide $b$ quark PDFs for calculating the production of bottom quarks
with high transverse momentum at LHC.  In a recent review
\cite{Thompson:2007mx}, HERA data for $F_2^c$ and $F_2^b$ are compared
with calculations using different schemes.  With further data from run
II of HERA, experimental errors are expected to decrease
significantly, and heavy flavor structure functions should provide
stringent tests of the theory description and the gluon density.


\subsection{Recent parton density fits}

The progress in the determination of PDFs is documented in a number of
recent fits, of which I can only mention a few:
\begin{itemize}
\item The MRST 2006 parton set \cite{Martin:2007bv} updates previous
  analyses by the MRST group.  It is a global fit to data for numerous
  processes, with inclusive DIS and Drell-Yan lepton pair production
  calculated at NNLO (other processes like jet production are only
  available at NLO).  Improvements in describing the charm and bottom
  contributions to $F_2$ at NNLO have led to significant changes
  compared with the MRST 2004 partons.  These changes increase the $W$
  and $Z$ production cross sections at LHC by about $6\%$.
\item The CTEQ6.5 partons \cite{Lai:2007dq,Pumplin:2007wg} are
  obtained from a global fit to data using NLO theory.  Significant
  changes are obtained compared with the previous CTEQ6.1 set, which
  used a ZM-VFNS whereas the new fit employs a GM-VFNS, which is much
  more adequate for describing the heavy flavor part of the HERA
  structure functions.  The influence of these changes on key
  processes at LHC is clearly seen in Fig.~\ref{fig:xsec}.
\item The Alekhin 06 parameterization \cite{Alekhin:2006zm} is
  obtained from only DIS and Drell-Yan data, which allows one to use
  NNLO accuracy throughout.  The heavy-flavor contributions to $F_2$
  are treated at order $\alpha_s^2$ in the FFNS with $n_f=3$.
\item The analysis in \cite{Blumlein:2006be} is restricted to DIS data
  and to non-singlet combinations of PDFs.  This avoids uncertainties
  on the gluon density and allows for fits at NNLO and NNNLO accuracy.
\end{itemize}
Apart from extracting parton densities, such analyses also permit an
extraction of the strong coupling constant $\alpha_s$, whose precision
has become very competitive with other determinations, see
\cite{Blumlein:2006be,Blumlein:2007dk}.  In this field, HERA
contributes both with DIS structure functions in a wide range of $Q^2$
and with dedicated analyses of jet production \cite{Glasman:2007sm}.


\subsection{High $Q^2$}

\begin{figure*}
\parbox{0.65\textwidth}{
\includegraphics[width=0.6\textwidth,%
  bb=26 72 531 403,clip=true]{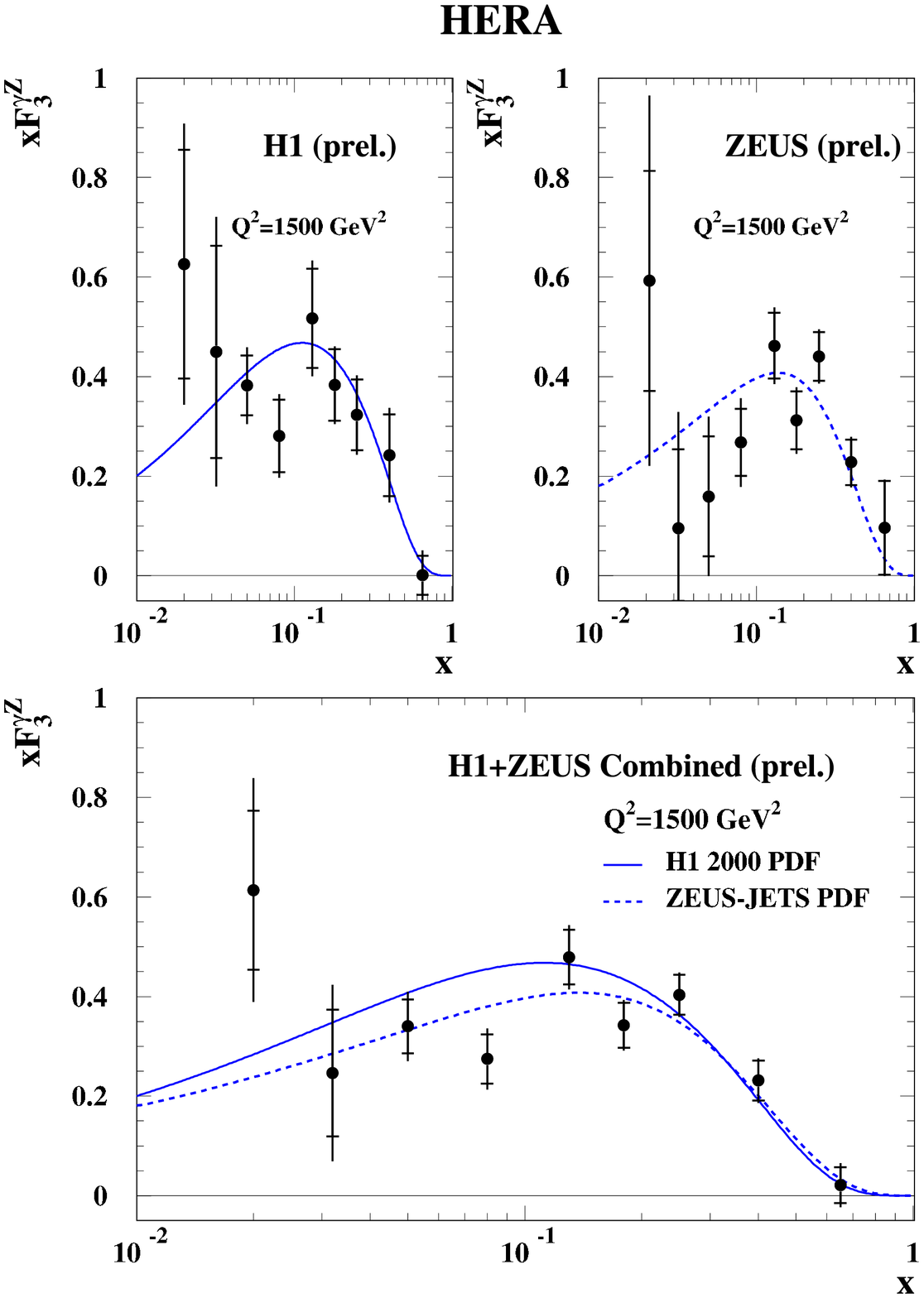}
}
\parbox{0.34\textwidth}{
  \caption{\label{fig:F3} HERA results for the structure function
    $F_3^{\gamma Z}$, which describes the interference of $\gamma$ and
    $Z$ exchange \protect\cite{hera-F3}.  The data shown in this plot
    correspond to about half of the full HERA statistics.}
}
\end{figure*}

For separating the densities of different quarks and antiquarks,
crucial information comes from experiments other than HERA.  In
particular, Drell-Yan lepton pairs and $W^\pm$ production at the
Tevatron \cite{Petroff:2007lp} provide a handle for the separation of
$\bar{u}$ and $\bar{d}$ distributions, whereas data from DIS
experiments with $\nu$ and $\bar{\nu}$ beams give the strongest
constraints on $s$ and $\bar{s}$ distributions \cite{Lai:2007dq}.
Nevertheless, structure function data from HERA contribute to this
field as well, in particular in the large $Q^2$ region, where $Z$ or
$W$ exchange becomes measurable.  The theoretical interpretation of
such data is very clean, since there are no nuclear corrections and
since inclusive DIS can be analyzed at NNLO.  Important observables
are the structure functions for the interference of $\gamma$ and $Z$
exchange: the lepton beam charge asymmetry gives access to
$F_3^{\gamma Z}$, which permits a clean measurement of the valence
quark combinations $q-\bar{q}$, and $F_2^{\gamma Z}$ from the lepton
beam polarization asymmetry is sensitive to the flavor combination
$u+d$.  Results for $F_3^{\gamma Z}$ are shown in Fig.~\ref{fig:F3}.
Charged current DIS in $e^+ p$ involves the combinations $u+c$ and
$\bar{d}+\bar{s}$, whereas its counterpart in $e^- p$ involves
$\bar{u}+\bar{c}$ and $d+s$.  It will be interesting to see whether a
selective measurement of the strange quark distribution by tagging
charm in the final state is practically feasible.  In all measurements
just mentioned, the size of experimental errors is crucial, and one
will have to wait for the analysis of the full HERA data set to assess
their impact on the precision of PDFs for different quark and
antiquark flavors.  Such detailed knowledge becomes important in
flavor sensitive new-physics channels at LHC, as has been illustrated
in \cite{Lai:2007dq} for the example of charged Higgs production via
$\bar{s}+c\to H^+$.


\subsection{Small $x$: higher orders}

It has long been known that higher perturbative orders play an
increasingly important role as $x$ becomes small.  This is for
instance illustrated by the drastic differences between the gluon
densities extracted from DIS data at LO, NLO, and NNLO accuracy.
These differences propagate to observables, notably to the structure
function $F_L(x,Q^2)$ for longitudinal photon polarization, see e.g.\
\cite{Thorne:2005qd}.  The reason for these effects are large
logarithms $\alpha_s^{n+m} \log^n(1/x)$ in the hard-scattering
coefficients and the evolution kernels for the parton densities.  It
is an ongoing program to sum these logarithms to all orders using the
BFKL formalism, see e.g.\
\cite{Forte:2006mk,White:2006yh,Ciafaloni:2007gf}.  Projecting out the
leading-twist part, i.e.\ the leading terms in $1/Q^2$, opens the
possibility to combine the resummed results at small $x$ with the
fixed-order ones at higher $x$.  The state-of-the art is
next-to-leading logarithmic accuracy, where terms $\alpha_s^{n+m}
\log^n(1/x)$ with $m=0,1$ are resummed, and progress has recently been
made on difficult technical issues such as the treatment of the
running coupling and the inclusion of quarks.  A first application in
a global parton density fit has been presented in \cite{White:2006yh},
with a very clear impact of resummation on both the gluon density and
on $F_L$.  It will be interesting to see the further development of
these efforts.

The longitudinal structure function $F_L$ is a basic observable in
DIS, at par with the well-measured structure function $F_2$.  Starting
at order $\alpha_s$, it is more directly sensitive to the gluon
distribution than $F_2$, and not surprisingly it turns out to
discriminate strongly between different theoretical treatments of the
gluon sector.  The low- and medium-energy runs of HERA in the last
months of its operation permit a direct measurement of $F_L$ by
Rosenbluth separation.  This will provide data in a region of moderate
$Q^2$ and small $x$ where the studies just mention show indeed
striking differences for various theory inputs.


\subsection{Small $x$: nonlinear effects}
\label{sec:sat}

So far we have discussed deep inelastic scattering in terms of the
leading-twist approximation, which is based on the limit of large
$Q^2$ and involves perturbatively calculable hard-scattering kernels
and parton distributions that evolve according to the DGLAP equations.
DGLAP evolution in particular describes the splitting process $g\to
gg$, in which gluons lose momentum, and which leads to a rapidly
growing gluon density at small $x$.  The transverse ``size'' of a
gluon as it is resolved in a hard process is given by the inverse
$1/Q$ of the large momentum scale.  When $g(x,Q^2)$ is so large that
gluons overlap in transverse space, they will recombine into gluons
with larger momentum, so that the growth of the density will be slowed
down and eventually saturate.  The region in $x$ and $Q^2$ where such
nonlinear effects become important is delineated by the saturation
scale $Q_s^2(x)$, which is a decreasing function of $x$.

\begin{figure}
\begin{center}
\includegraphics[width=0.38\textwidth]{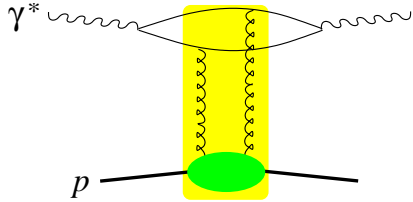}
\end{center}
\caption{\label{fig:dis-glu} The forward $\gamma^* p$ amplitude in the
  color dipole picture.  Its imaginary part gives the total $\gamma^*
  p$ cross section via the optical theorem.  The large shaded (yellow)
  box represents the dipole scattering amplitude $N_{q\bar{q}}$, and
  the lower (green) blob stands for the gluon distribution.}
\end{figure}

There are vigorous theoretical activities to describe this nonlinear
dynamics at various degrees of approximation.  I cannot describe them
within the limitations of this talk, and only mention keywords like
the Balitsky-Kovchegov equation, the JIMWLK equation, and pomeron
loops.  A recent overview and references can be found in
\cite{Iancu:2006qi}.  At the phenomenological level, nonlinear effects
in a large class of DIS processes can be described using the
color-dipole formulation.  Its basic ingredient is the amplitude
$N_{q\bar{q}}$ for the scattering of a quark-antiquark dipole on a
proton target, as shown in Fig.~\ref{fig:dis-glu}.  In the kinematic
region where nonlinear effects are unimportant, $N_{q\bar{q}}$ is
related to the gluon density, so that one can make contact with the
leading-twist description.  Several studies have proposed
phenomenological parameterizations of $N_{q\bar{q}}$ with saturation
behavior due to nonlinear effects, and fitted them to DIS data from
HERA.  This gives a good description of $F_2$ at small $x$, down to
$Q^2$ much below the values where the leading-twist description is
applicable.  More recently, due attention has also been paid to the
charm contribution $F_2^c$, where good agreement with the data is
achieved as well
\cite{Kowalski:2006hc,GolecBiernat:2006ba,Soyez:2007kg}.  This
description gains much credibility from the fact that with the same
nonperturbative input one can describe several diffractive processes,
as I will discuss in Section~\ref{sec:diff}.  The saturation mechanism
predicts geometric scaling, which states that $F_2$ at small $x$
depends on $x$ and $Q^2$ only in the combination $Q^2 /Q_s^2(x)$.
This scaling is indeed seen in HERA data, and I refer to
\cite{Avsar:2007ht} for a recent critical discussion and references.
Taken together, these features can be seen as strong indications that
nonlinear effects are at work in the small-$x$ region at HERA,
although there is no unambiguous proof of this so far.  It should also
be mentioned that the treatment of higher perturbative orders in the
theory of nonlinear effects is more difficult and much less advanced
than in the leading-twist approximation.

\begin{table}[b]
  \caption{\label{tab:sat} The saturation scale $Q_s(x)$
    obtained by recent analyses of HERA data in the dipole approach.}
\begin{center}
\renewcommand{\arraystretch}{1.1}
\begin{tabular}{ccc} \hline
$x=10^{-4}$ & $x=10^{-6}$ & Reference \\ \hline
$0.8~\gev$  & $4.0~\gev$  & \protect\cite{Kowalski:2006hc} \\
$0.8~\gev$  & $2.0~\gev$  & \protect\cite{GolecBiernat:2006ba} \\
$0.7~\gev$  & $1.9~\gev$  & \protect\cite{Soyez:2007kg} \\ \hline
\end{tabular}
\end{center}
\end{table}

Table~\ref{tab:sat} compares the saturation scales obtained in recent
phenomenological analyses.  For $x= 10^{-4}$, which is a typical value
for HERA, one finds $Q_s$ slightly below $1 \gev$.  This is at the
borderline of applicability for QCD perturbation theory, which is part
of the difficulty to prove or disprove the validity of the theoretical
approach in the HERA regime.  The spread in $Q_s$ at $x= 10^{-6}$ is
hardly surprising and indicates the theoretical uncertainties involved
in such an extra\-polation.  HERA measurements have driven many
efforts to quantify the onset of nonlinear effects in hard scattering
at high energy.  With detectors for very forward particles at LHC,
there is the prospect of pursuing such studies in $pp$ collisions.  It
has for instance been estimated \cite{Albrow:2006xt} that one can
detect Drell-Yan lepton pairs with $x\lsim 10^{-6}$ and invariant mass
$Q^2\gsim 4 \gev^2$ using the CASTOR calorimeter.  According to
Table~\ref{tab:sat}, nonlinear effects may well be strong in that
region.  The study of saturation effects at HERA has also helped in
developing a quantitative description of the initial state in
heavy-ion collisions, presently at RHIC and soon at ALICE, see e.g.\
\cite{Venugopalan:2007vb}.

Expanding the DIS cross section obtained in the dipole approach in
powers of $1/Q^2$, one can isolate the leading term, which corresponds
to what is described by the leading-twist approach.  An early study in
\cite{Bartels:2000hv} found for the ratio of the full result and its
twist-two approximation
\begin{align}
  \label{sat-est}
\frac{F_2^{\text{full}}}{F_2^{t=2}} &\approx 0.94 \,, &
\frac{F_L^{\text{full}}}{F_L^{t=2}} &\approx 0.66 \quad
\end{align}
at $Q^2= 5 \gev^2$ and $x= 2.5\times 10^{-4}$, which is a typical
point in HERA kinematics.  We see that the longitudinal structure
function is much more affected than $F_2$ by nonlinear effects, which
highlights once more the dynamical sensitivity of this observable.
Unfortunately, the numbers in eq.~\eqref{sat-est} are based on a
fairly old parameterization of the dipole scattering amplitude.  An
update of such a study for more recent models would be useful, since a
result as in eq.~\eqref{sat-est} would have important consequences on
the theoretical error one should assign to extractions of PDFs using
the leading-twist approximation of $F_2$.


\section{Diffraction}
\label{sec:diff}

A striking discovery of HERA is that a large fraction (around 10\%) of
events in DIS have a leading proton or a large rapidity gap between
the proton remnant and the other hadrons in the final state.  The
increasing precision and kinematic coverage of the data
\cite{Aktas:2006hy,ZEUS:2007ep} reveals in particular two crucial
features of the cross section for inclusive diffraction, $\gamma^*
p\to X +p$.
\begin{figure*}[t]
\hspace{0.5em}
\includegraphics[width=0.43\textwidth]{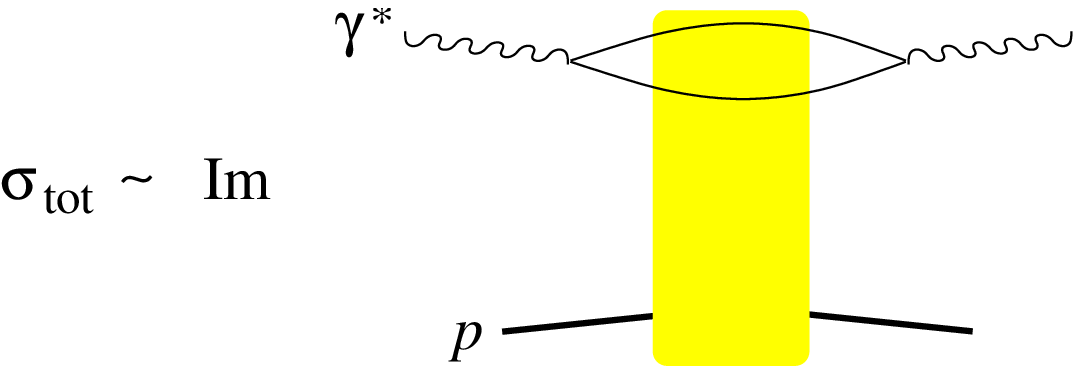}
\hspace{4em}
\includegraphics[width=0.38\textwidth]{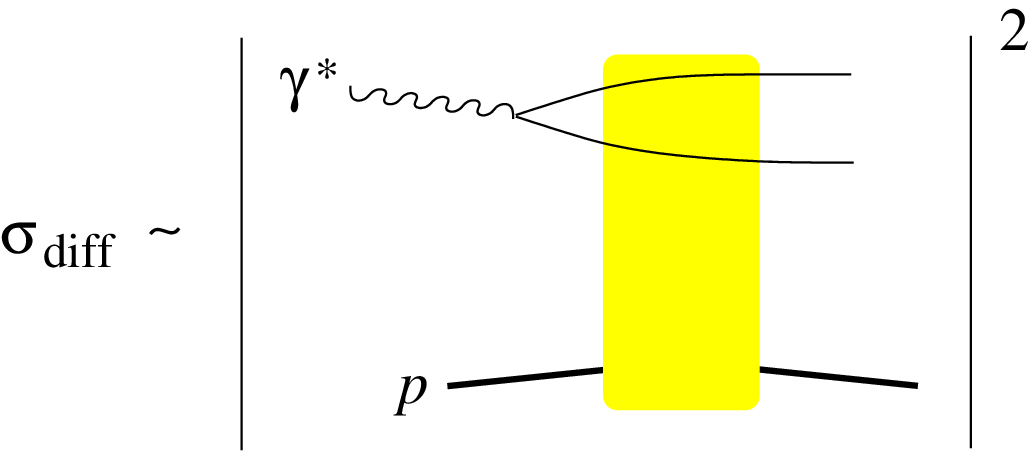}
\caption{\label{fig:dis-diff} Description of the total and of the
  diffractive $\gamma^* p$ cross section in the dipole approach at
  lowest order in $\alpha_s$.  The shaded (yellow) boxes represent the
  dipole scattering amplitude.}
\end{figure*}
\begin{itemize}
\item The ratio $\sigma_{\text{diff}}/ \sigma_{\text{tot}}$ of the
  diffractive and the total $\gamma^* p$ cross sections is rather flat
  in $Q^2$ for fixed $x$ (except in corners of phase space).  This
  means that inclusive diffraction is a leading-twist phenomenon in
  $\gamma^* p$ interactions and gives a significant contribution to
  the inclusive structure function $F_2$ at high $Q^2$.

  This observation contains an important general lesson.  There is no
  doubt that the leading-twist description of the \emph{inclusive}
  $\gamma^* p$ cross section is adequate at high $Q^2$: it is based on
  factorization theorems in QCD and works to high precision in
  practice.  This does however not imply that the same type of
  leading-twist description, together with standard hadronization
  models as they are implemented in event generators, gives an
  adequate description of the \emph{final state}.  Such a description
  fails to account for the large observed fraction of rapidity gap
  events.  {}From the theory side this not surprising: the derivation
  of factorization theorems for inclusive observables heavily relies
  on taking a \emph{sum} over final states.
\item At given $Q^2$ the cross section ratio $\sigma_{\text{diff}}/
  \sigma_{\text{tot}}$ has a very flat dependence on the collision
  energy, or equivalently on $x$.  As indicated in
  Fig.~\ref{fig:dis-diff}, both cross sections can be calculated in
  the dipole approach.  The saturation mechanism discussed in
  Section~\ref{sec:sat} provides a natural explanation of this
  surprising result, which can be understood using simple analytical
  approximations and is numerically seen to high precision
  \cite{GolecBiernat:1999qd}.  It should be mentioned that in a
  restricted kinematic region, models where the dipole scattering
  amplitude does not exhibit saturation can also describe the data
  \cite{Forshaw:2006np}.  This underscores the importance of having
  precise measurements in a wide kinematic region.
\end{itemize}
Compared with inclusive diffraction, the cross sections for exclusive
diffractive channels such as vector meson production ($\gamma^* p\to V
p$) or virtual Compton scattering ($\gamma^* p\to \gamma p$) at high
$Q^2$ grow significantly faster with energy.  This can be understood
generically in the saturation approach, and dipole models such as the
one in \cite{Kowalski:2006hc} provide a good quantitative description
of these channels as well.  The relevant graph for $\gamma^* p\to
\gamma p$ has the same form as in Fig.~\ref{fig:dis-glu}.  In the
region of large $Q^2$, where nonlinear effects are negligible, these
exclusive processes also admit a leading-twist description, where the
gluon density appearing in inclusive processes is replaced by the
generalized gluon distribution, which depends on the difference of
proton momenta in the initial and final state.  Its dependence on the
invariant momentum transfer $t$ contains information about the spatial
distribution of gluons in the impact parameter plane, whereas $x$
determines the relevant longitudinal momentum fractions
\cite{Diehl:2002he}.  Exclusive diffractive processes thus open the
possibility to obtain a three-dimensional picture of how gluons and
sea quarks are distributed in the proton.  A recent quantitative study
can be found in \cite{Kumericki:2007sa}, and the relevance of such
information for physics at LHC has been discussed in
\cite{Frankfurt:2003td}.


\subsection{$pp$ and $p\bar{p}$ collisions}

\begin{figure*}[t]
\begin{center}
\includegraphics[width=0.36\textwidth]{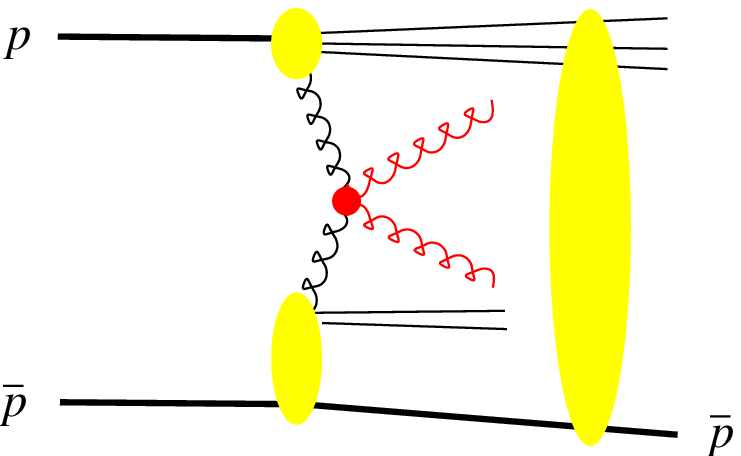}
\hspace{4em}
\includegraphics[width=0.36\textwidth]{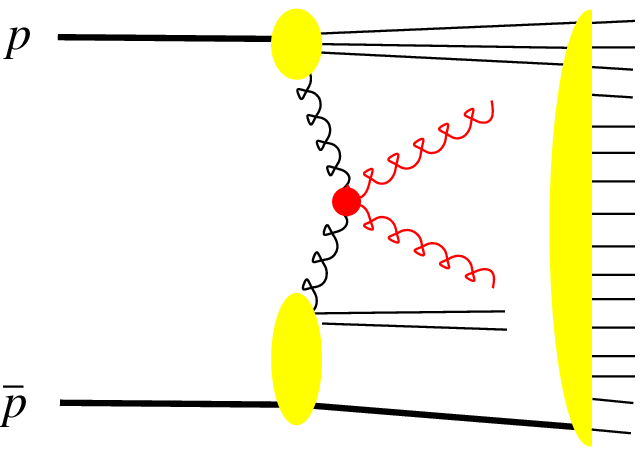}
\end{center}
\caption{\label{fig:cdf} Left: Dijet production with a leading
  antiproton, $p\bar{p}\to \text{dijet} + X + \bar{p}$. The small
  upper blob represents the usual gluon density, the small lower blob
  stands for the diffractive gluon density, and the large blob denotes
  interactions between spectator partons in the proton and antiproton.
  Right: The same type of spectator interactions can populate the
  final state with particles, thus destroying the rapidity gap.}
\end{figure*}

A key for the successful QCD description of diffraction at HERA is
that the presence of a hard scale such as $Q^2$ allows us to calculate
part of the process in perturbation theory.  It is natural to attempt
the same in diffractive processes at $pp$ and $p\bar{p}$ colliders.
The CDF Collaboration has studied a number of channels with a leading
antiproton or a large rapidity gap, and with a hard scale in the final
state provided e.g.\ by jets, weak gauge bosons or heavy quarks
\cite{Goulianos:2004as}.  The fraction of rapidity gaps in such hard
processes is found to be at the percent level, about an order of
magnitude smaller than the fraction of rapidity gaps in DIS.
If one assumes factorization and calculates e.g.\ diffractive jet
production, $p\bar{p}\to \text{dijet} + X + \bar{p}$ in terms of
diffractive parton densities extracted from HERA DIS data, one obtains
a rate substantially larger than what is measured
\cite{Affolder:2000vb}.  This indicates that in diffractive $p\bar{p}$
and $pp$ interactions, factorization into parton densities and a
partonic hard-scattering subprocess is strongly broken.  The analysis
of factorization proofs indeed shows that interactions between the
spectator partons of both hadrons cannot be neglected in such a
situation: their effects only cancel in sufficiently \emph{inclusive}
observables, where no rapidity gap is required \cite{Collins:2001ga}.
Such interactions ``bypass'' the hard scattering and are at least
partly soft, so that one has to resort to phenomenological models in
order to describe their effect.  The combination of diffractive data
from HERA and the Tevatron allows one to test such models.

Spectator interactions can also populate the final state with
additional particles and thus destroy any rapidity gap, as shown in
Fig.~\ref{fig:cdf}.  There is hence a link between the suppression of
hard diffraction and the physics of multiple interactions and the
underlying event.  The latter is expected to be of great importance at
LHC, given the huge phase space available for producing energetic
particles by secondary interactions in a single $pp$ collision.


\subsection{Exclusive diffraction at LHC}

\begin{figure}
\begin{center}
\includegraphics[width=0.34\textwidth]{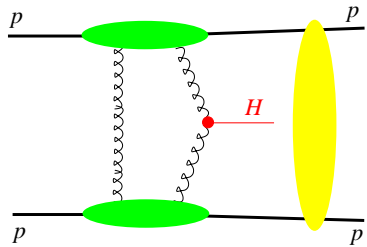}
\end{center}
\caption{\label{fig:higgs} Schematic representation of exclusive Higgs
  production, $pp\to p+H+p$.  The smaller (green) blobs represent the
  generalized gluon distribution, and the larger (yellow) one stands
  for spectator interactions.}
\end{figure}

A particular type of diffractive events at LHC has received
considerable attention.  If both protons are scattered diffractively,
one can produce new particles like the Higgs in a very clean
environment, $pp\to p+H+p$.  The same holds for any other particle
with a sufficiently large coupling to two gluons.  To measure such
events is an experimental challenge, requiring very forward detectors
and having to overcome relatively low rates and trigger issues, and
there are ongoing efforts to achieve this goal
\cite{Albrow:2005ig,Albrow:2006xt}.  Advantages of exclusive
production are the possibility to measure mass and width to an
accuracy of a few GeV, the determination of quantum numbers (since
$CP=++$ states are strongly enhanced by the production mechanism), and
a generally very favorable signal-to-background ratio.  The
calculation of the cross section is nontrivial and involves a number
of ingredients.  Among these are in particular the generalized gluon
distribution, which can be extracted from exclusive diffractive
channels at HERA, radiative corrections in the form of Sudakov
factors, as well as spectator interactions as discussed in the
previous subsection.  A simplified graph is shown in
Fig.~\ref{fig:higgs}.  Corresponding theoretical calculations are in
fair agreement with measurements of similar exclusive channels at the
Tevatron, such as dijet or $\gamma\gamma$ production
\cite{Goulianos:2007xr}.

Detailed experimental studies have been performed for the production
of a light standard model Higgs in the decay channel $H\to W^+ W^-$,
with one $W$ being off-shell for $m_H < 2 m_W$.  For an integrated
luminosity of $30 \operatorname{fb}^{-1}$, a few signal events after
triggers and cuts are expected according to \cite{Cox:2005if}, with a
very low background.  The decay channel $H\to b\bar{b}$ is more
difficult experimentally, and for a standard model Higgs it will
probably yield too few events after triggers and cuts
\cite{Cox:2007sw}.  In a number of supersymmetric scenarios, however,
the production rate is significantly enhanced, especially for large
$\tan\beta$.  In such cases, exclusive Higgs production in $pp\to
p+b\bar{b}+p$ may be measurable with $3\sigma$ significance, and in
certain scenarios a $5\sigma$ discovery may even be possible with
sufficient running time \cite{Cox:2007sw,Heinemeyer:2007tu}.


\section{Conclusions}
\label{sec:con}

HERA has pioneered the study of deep inelastic scattering at small
Bjorken-$x$ and at large $Q^2$.  The kinematic coverage and precision
of the data for the inclusive structure function $F_2(x,Q^2)$ provides
a key input for the precise determination of parton densities, which
is a crucial prerequisite for calculating precise cross sections at
LHC.  The continuous improvement of the data has been accompanied by
substantial progress in the understanding of QCD dynamics at high
energy, from the calculation of perturbative corrections and their
resummation to the development of a theory for nonlinear effects at
very small $x$.

Diffraction in DIS exhibits a number of simple features, and at the
same time shows the intricacies of describing final states in QCD,
having invalidated a number of (too) naive expectations.  Comparison
of diffractive measurements at HERA and the Tevatron reveals the yet
higher complexity of the final state in hadron-hadron collisions.
Calculations based on these measurements can be used to estimate the
exclusive diffractive production of the Higgs and other new particles
at LHC.  While their detection requires a substantial experimental
effort, these clean final states can provide a valuable tool for
precise measurements in the Higgs sector, and in some scenarios of
physics beyond the Standard Model may even become a discovery channel.

Important results can be expected from the final analysis of the data
of HERA run II, since statistics and kinematic reach are still
limiting factors in important measurements, such as DIS at high $Q^2$,
the production of heavy flavors, or various diffractive channels.  The
direct measurement of the longitudinal structure function $F_L$ will
provide a missing cornerstone in the description of inclusive DIS and
a sharp discriminator between theoretical approaches to QCD at small
$x$ and moderate~$Q^2$. \\


\section*{Acknowledgments}

I wish to thank D. Son and his colleagues for organizing a wonderful
conference.  It is a pleasure to thank H. Abramowicz, J. Bartels, O.
Behnke, A. Cooper-Sarkar, L.  Dixon, C. Gwenlan, H. Jung, M. Klein, G.
Kramer, S. Moch and L. Motyka for discussions and valuable input.
Special thanks go to J. Bartels for a careful reading of the
manuscript.


\end{document}